\documentclass{emulateapj}

\received{2015 January 30}
\slugcomment{submitted to {\it the Astrophysical Journal, Letter}}

\shortauthors{KOBAYASHI et al.}
\shorttitle{Subclasses of Type Ia Supernovae as the origin of [$\alpha$/Fe] ratios in dwarf spheroidal galaxies}

\def\gtsim {>\kern-1.2em\lower1.1ex\hbox{$\sim$}~}   
\def\ltsim {<\kern-1.2em\lower1.1ex\hbox{$\sim$}~}   

\begin{document}

\title{Subclasses of Type Ia Supernovae as the origin of [$\alpha$/Fe] ratios in dwarf spheroidal galaxies}
\author{Chiaki KOBAYASHI$^{1,2,3}$, Ken'ichi NOMOTO$^{3,}$\altaffilmark{*}, and Izumi Hachisu$^4$}
\affil{$^1$ School of Physics, Astronomy and Mathematics,
Centre for Astrophysics Research, University of Hertfordshire,
College Lane, Hatfield AL10 9AB, UK; c.kobayashi@herts.ac.uk}
\affil{$^2$ Distinguished Visitor, Research School of Astronomy \& Astrophysics, The Australian National University, Cotter Rd., Weston ACT 2611, Australia}
\affil{$^3$ Kavli Institute for the Physics and Mathematics of the Universe (WPI), The University of Tokyo, Kashiwa, Chiba 277-8583, Japan}
\altaffiltext{*}{Hamamatsu Professor}
\affil{$^4$ Department of Earth Science and Astronomy, 
College of Arts and Sciences, The University of Tokyo, 
Meguro-ku, Tokyo 153-8902, Japan}

\begin{abstract}
Recent extensive
observations of Type Ia Supernovae (SNe Ia) have revealed the existence of a diversity of SNe Ia, including SNe Iax.
We introduce two possible channels in the single degenerate scenario:
1) double detonations in sub-Chandrasekhar (Ch) mass CO white dwarfs (WDs),
where a thin He envelope is developed with relatively low accretion rates after He novae even at low metallicities,
and
2) carbon deflagrations in Ch-mass {\sl possibly} hybrid C+O+Ne WDs,
where WD winds occur at [Fe/H] $\sim -2.5$ at high accretion rates.
These subclasses of SNe Ia are rarer than `normal' SNe Ia and do not affect the chemical evolution in the solar neighborhood, but can be very important in metal-poor systems with stochastic star formation.
In dwarf spheroidal galaxies in the Local Group, the decrease of [$\alpha$/Fe] ratios at [Fe/H] $\sim -2$ to $-1.5$ can be produced depending on the star formation history.
SNe Iax give high [Mn/Fe], while sub-Ch-mass SNe Ia give low [Mn/Fe], and thus a model including a mix of the two is favoured by the available observations.
\end{abstract}


\section{Introduction}

Elemental abundance ratios can be used as a `cosmic clock' because different elements are produced from stars on different time-scales.
Stars more massive than $\sim 8M_\odot$ explode as core-collapse supernovae, which produce more $\alpha$ elements than Fe. Among core-collapse supernovae, [$\alpha$/Fe] ratios are larger for more massive progenitors, and the [$\alpha$/Fe] ratios weighted with the initial mass function (IMF) are $\sim +0.4$, which are consistent with the plateau values of observed [$\alpha$/Fe] ratios of metal-poor stars in the solar neighborhood \citep[e.g.,][]{cay04}.
On the other hand, thermonuclear supernovae, i.e., Type Ia Supernovae (SNe Ia), produce more iron-peak elements, in particular Fe and Mn. This results in the decreasing trend of [$\alpha$/Fe] and the increasing trend of [Mn/Fe] from [Fe/H] $\sim -1$ to $\sim 0$ in the solar neighborhood \citep{gra89,kob06,kob11agb}.

The progenitors of SNe Ia are still a matter of debate between (1) deflagrations or delayed detonations of Chandrasekhar (Ch) mass white dwarfs (WDs) from single degenerate systems, (2) sub-Ch-mass explosions from double degenerate systems, or (3) double detonation of sub-Ch-mass WDs in single or double degenerate systems \citep[e.g.,][]{hil00}.
The nucleosynthesis yields that are most commonly used in galactic chemical evolution (GCE) models are for the W7 model \citep{nom97}, which is a deflagration of a Ch-mass WD. This model gives good agreement with the observed [$\alpha$/Fe] and [Mn/Fe] ratios, in combination with our metallicity-dependent SN Ia progenitor model \citep{kob09}.
\citet{sei13} showed that sub-Ch models result in too low Mn abundances.
Mn (decayed from $^{55}$Co) is produced from nuclear statistical equilibrium (NSE) in the center and incomplete Si-burning in the outer region, and the yield from the former process is much reduced in sub-Ch models because of the lower density \citep{shi92}.

The lifetime, or delay time, of SNe Ia has been estimated from the metallicity of the `knee' in the [$\alpha$/Fe]-[Fe/H] relations as $\sim 1.5$Gyr.
However, the lifetime/delay-time distribution function that is estimated from observed SN Ia rates or is calculated from binary population synthesis models spans a wide range from $\sim 0.01$Gyr to $\sim 20$Gyr with a peak at $\sim 0.1$Gyr \citep[e.g.,][]{mao14}. This is too short to reproduce the [$\alpha$/Fe] knee at [Fe/H] $\sim -1$ in the solar neighborhood.
However, \citet{kob98} showed that in single degenerate systems, the SN Ia lifetime depends on the metallicity of progenitor systems because of WD winds \citep{hac96,hac12}, and with this metallicity effect, succeeded in reproducing the knee at [Fe/H] $\sim -1$ in the solar neighborhood.

This metallicity effect inhibits the enrichment from SNe Ia in very low-metallicity systems. This is consistent with observed [$\alpha$/Fe] ratios of globular cluster systems \citep[e.g.,][]{car09} where asymptotic giant branch (AGB) stars (with longer lifetimes than SNe Ia) contribute to the O-Na anti-correlation and/or s-process abundances \citep[e.g.,][]{kra97}.
However, the metallicity effect may conflict with the observations of dwarf spheroidal (dSph) galaxies that now show a quite clear knee at lower metallicities; [Fe/H] $\sim -2$ in Sculptor, $\sim -1.5$ in Fornax, and so on \citep[e.g.,][]{tol09}.
If this knee is caused by SNe Ia, [Mn/Fe] ratios should show an increasing trend with [Fe/H] from this [$\alpha$/Fe] knee. However, in dSph galaxies, [Mn/Fe] ratios are as low as in the Galactic halo stars for a wide range of metallicity \citep{mcw03,rom11,nor12}.
One possible scenario is the lack of contributions from massive core-collapse supernovae due to the incomplete sampling of IMF \citep{tol03,koc08,ven12,nom13}. Less-massive supernovae ($\sim 20M_\odot$) give low [$\alpha$/Fe] ratios without changing [Mn/Fe] ratios.

In this Letter, we propose a new scenario that includes new subclasses of SNe Ia that have been discovered with recent extensive observations of supernovae \citep[e.g.,][]{li11,sca14}.
We introduce two possible channels for such SNe Ia (\S 2) and apply these channels to our GCE models of the solar neighborhood and dSph galaxies (\S 3).
Our conclusions are presented in \S4.

\section{Progenitor Model}

\subsection{Sub-Chandrasekhar mass SNe Ia}

\begin{figure}
\center 
\includegraphics[width=8.2cm]{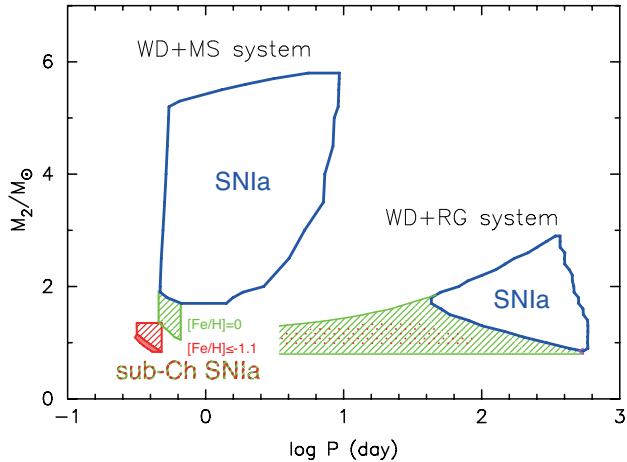}
\caption{\label{fig:sch}
Regions for sub-Ch SNe Ia (shaded and hatched area) in the diagram of initial orbital period $P$ vs. initial secondary mass $M_2$ at $Z=0.001$ (red) and $Z=Z_\odot$ (green), where the He accretion rate lies in $\dot{M}=3-4\times10^{-8}M_\odot{\rm yr}^{-1}$.
For $Z=0.001$, the WD+RG system appears only with the initial WD mass $M_{\rm WD,0}=1.2 M_\odot$ (dotted area).
All other regions are for $M_{\rm WD,0}=1.0 M_\odot$.
For comparison, the regions for SNe Ia at $Z=Z_\odot$ (blue) are taken from \citet{hac12}, where the WD mass reaches Ch-mass with the wind and stripping effects ($c_1=3$).
}
\end{figure}

This type of explosion is expected to occur when the He layer grows at 
a relatively slow rate ($<4\times10^{-8} M_\odot {\rm yr}^{-1}$; \citealt{nom82}, see also \citealt{ibe91}).
Previous models showed that He detonations produce copious $^{56}$Ni 
near the surface in the He layer, so that
the He features are too strong and the envelopes are too hot 
to be consistent with observed SN Ia spectra \citep{hof96,nug97}.
Thus, it was thought that such a double detonation may not occur.

However, recent studies \citep[e.g.,][]{she09,fin10,woo11}
showed that a thin He layer containing
less than $0.04 M_\odot$ may produce an He detonation to trigger a double
detonation, while the mass of the He layer is too small to produce 
a significant amount of $^{56}$Ni and prominent He features.
This small He mass for the He detonation makes these double-detonation sub-Ch-mass models viable progenitors of SNe Ia,
as the light curves and spectra are consistent with some observations \citep[e.g.,][]{sim10,kro10}.

\begin{figure}
\center 
\includegraphics[width=8.7cm]{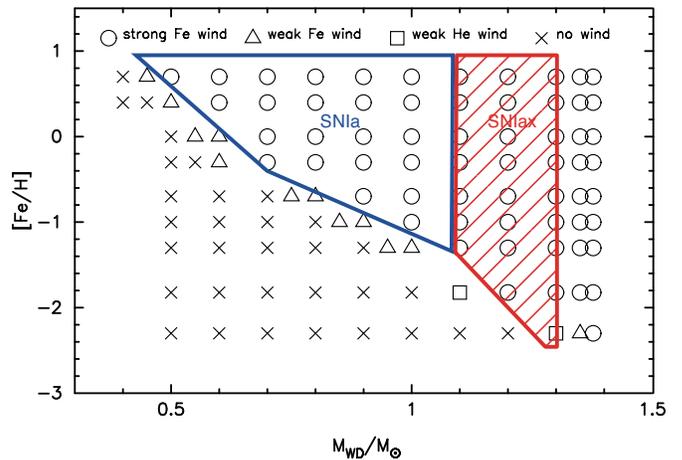}
\caption{\label{fig:iax}
The region for SNe Iax (hybrid C+O+Ne WDs, red hatched area) in the diagram of WD mass vs. iron abundance, comparing with that for SNe Ia (blue area).
This is similar to Fig.1 of \citet{kob98}, but the wind solutions are updated for a few models by \citet{hac12}.
The symbols indicate strong winds (circles), weak Fe winds where the wind velocity does not exceed the escape velocity (triangles), weak He winds that are driven not by Fe lines but by He lines (squares), and no winds (crosses).
}
\end{figure}

\begin{table*}[t]
\begin{center}
\begin{tabular}{lccccccccc}
\hline
[Fe/H] & -$\infty$ & -2.5 & -2.4 & -2 & -1.1 & -1 & -0.7 & 0. & 0.4\\
\hline
\multicolumn{10}{c}{normal SN Ia}\\
\hline
$m_{{\rm RG},\ell}$ & -& -& -& -& 0.9 & 0.9 & 0.9 & 0.9 & 0.8\\
$m_{{\rm RG,u}}$   & -& -& -& -& 0.9 & 1.5 & 2.0 & 3.0 & 3.5\\
$m_{{\rm MS},\ell}$ & -& -& -& -& 1.8 & 1.8 & 1.8 & 1.8 & 1.8\\
$m_{{\rm MS,u}}$   & -& -& -& -& 1.8 & 2.6 & 4.0 & 5.5 & 6.0\\
$m_{{\rm WD},\ell}$ & -& -& -& -& 2.4 & 2.5 & 2.8 & 3.5 & 3.9\\
$m_{\rm WD,u}$    & -& -& -& -& 6.9 & 7.0 & 7.3 & 8.0 & 8.4\\
\hline
\multicolumn{10}{c}{sub-Ch SN Ia}\\
\hline
$m_{{\rm subCh,RG},\ell}$ & -& -& -& -& 0.835 & 0.835 & 0.835 & 0.835 &\\
$m_{{\rm subCh,RG,u}}$   & -& -& -& -& 0.835 & 1.0 & 1.3 & 1.9 &\\
$m_{{\rm subCh,MS},\ell}$ & 0.835 & & & & 0.835 & 1.05 & 1.05 & 1.05 &\\
$m_{{\rm subCh,MS,u}}$   & 1.35 & & & & 1.35 & 1.9 & 1.9 & 1.9 &\\
$m_{{\rm WD},\ell}$       & 5.9 & & & & 5.9 & 6.0 & 6.3 & 7.0 &\\
$m_{\rm WD,u}$           & 6.9 & & & & 6.9 & 7.0 & 7.3 & 8.0 &\\
\hline
\multicolumn{10}{c}{SN Iax}\\
\hline
$m_{{\rm Iax,RG},\ell}$ & -& 0.8 & 0.8 & 0.8 & & 0.8 & & 0.8 &\\
$m_{{\rm Iax,RG,u}}$   & -& 0.8 & 1.5 & 3.0 & & 3.0 & & 3.0 &\\
$m_{{\rm Iax,MS},\ell}$ & -& 1.6 & 1.6 & 1.6 & & 1.6 & & 1.6 &\\
$m_{{\rm Iax,MS,u}}$   & -& 1.6 & 2.9 & 6.5 & & 6.5 & & 6.5 &\\
$m_{{\rm WD},\ell}$ & -& 6.3 & 6.34 & 6.5 & & 7.0 & & 8.0 &\\
$m_{\rm WD,u}$     & -& 7.3 & 7.34 & 7.5 & & 8.0 & & 9.0 &\\
\hline
\end{tabular}
\caption{\rm
Initial mass ranges of primary stars, and red-giant and main-sequence companions, as a function of iron abundance.
The dash entries indicate no model, while interpolation is applied in the GCE code for the blank entries.
}
\end{center}
\end{table*}

Such a thin He layer can be formed by the accretion from H-rich donors \citep[e.g.,][]{yun95} or from He-WDs or He-burning stars \citep[e.g.,][]{rui14}, if the He accretion rate lies in a narrow range of $\dot{M} \sim 3-4\times10^{-8} M_\odot {\rm yr}^{-1}$.
In this Letter, we assume that the thin He layer ($\le 0.04M_\odot$) is formed by H-rich donors, which detonates at the ignition density of $\sim 2\times10^6 {\rm g\,cm}^{-3}$ \citep[e.g.,][]{nom82}.
The required $\dot{M}$ slightly depends on the WD mass \citep{kaw87}, but is always lower than for Ch-mass explosions.
Therefore, in the diagram of secondary mass and orbital period (Fig.\ref{fig:sch}), the sub-Ch SNe Ia regions are located below those for Ch-mass SNe Ia.
At [Fe/H] $\ltsim -1.1$,
because WD winds are too weak to eject H-rich envelopes, 
the processed He effectively
accumulates on to the WDs in every H-shell flash of novae.
Assuming this He efficiency of 50\% \citep{hac01},
the sub-Ch region (shaded area) is determined from
the initial accretion rates, which depend on
the thermal timescales for main-sequence (MS) companions.
At larger $P$, the stellar radius becomes larger,
and thus a smaller secondary star gives the required $\dot{M}$.
This results in the downward slope in this diagram.
With a slightly larger initial $\dot{M}$ (hatched area), the systems undergo He nova explosions, but will cause delayed He detonations when $\dot{M}$ drops.
Similar $\dot{M}$ regions appear just below the normal SN Ia regions, also with strong WD winds at [Fe/H] $\gtsim -1.1$, both for MS and red-giant (RG) companions.
For the RG companions, however, if there is no wind, the systems with $M_2/M_{\rm WD,0}\ge 0.79$ cause a common envelope \citep{hac96}, the parameter space for which is limited only for hybrid WDs (\S 2.2) with $M_{\rm WD,0} \gtsim 1.1M_\odot$ (dotted area), and the contribution is neglected in this Letter.

The initial mass ranges for the secondary and primary stars in our GCE models are summarized in Table 1.
For the primary stars, the WD should be relatively massive, $M_{\rm WD,0} \sim 1.0-1.1 M_\odot$, because for lower-mass WDs, the He mass required for the ignition is larger \citep{she09,woo11} and the He features are more prominent.
According to stellar evolution models, the mass of C+O WDs is larger for larger mass and higher metallicity progenitors.
The primary mass ranges are set at the massive end of C+O WDs for normal SNe Ia ($M_{\rm WD,0} \sim 0.6-1.1M_\odot$).
Note that the primary mass range was constant, $3-8M_\odot$ independent of metallicity in \citet{kob09}, but depends on metallicity in this work, taken from \citet{ume99}.

The nucleosynthesis yields are taken from the $1.05M_\odot$ model of \citet{shi92};
$M({\rm Fe})=0.5643M_\odot, M({\rm O})=0.0594M_\odot$, $M({\rm Mn})=3.246 \times 10^{-3} M_\odot$, which gives [O/Fe] $=-1.86$ and [Mn/Fe] $=-0.26$
\footnote{\citet{and89} is used for the solar abundance as in our GCE \citep[e.g.,][]{kob11agb}.}.
This is very similar to the $1.06M_\odot$ model in \citet{sim10} and \citet{sei13}.
For sub-Ch models, Mn is mostly synthesised in incomplete Si-burning, and therefore the Mn yields should depend on metallicity. We include this effect as $M({\rm Mn}) \propto Z^{3}$ \citep{sei15,yam15}.

\subsection{2002cx-like SNe Ia}

\begin{figure*}
\center 
\includegraphics[width=15cm]{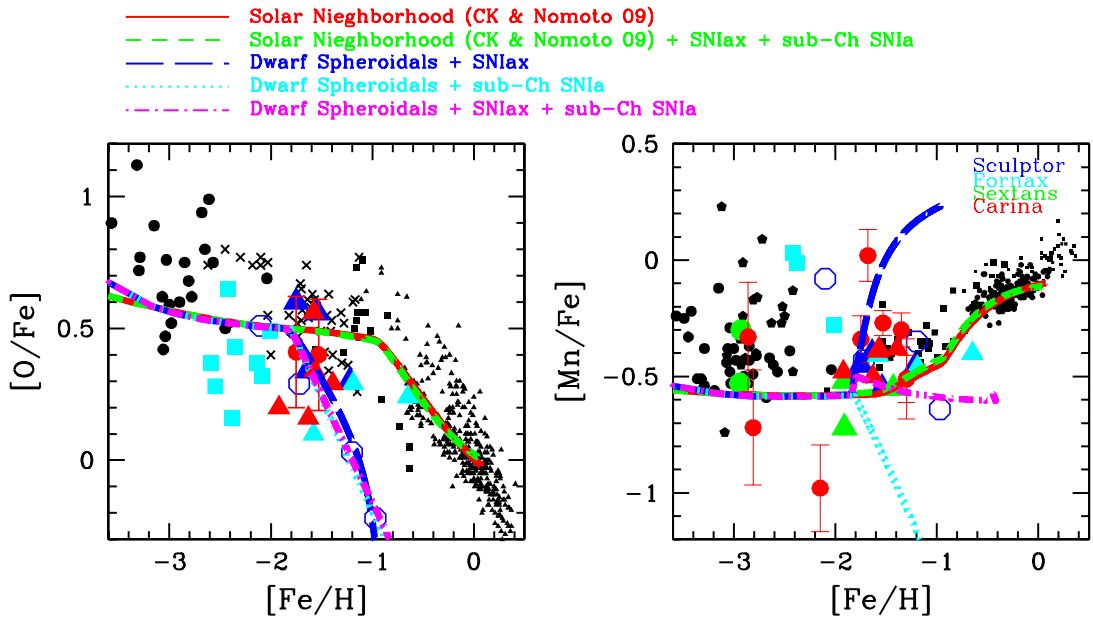}
\caption{\label{fig:xfe}
Evolution of elemental abundance ratios for the solar neighborhood with only normal SNe Ia (red solid lines) and a model with 50\% SNe Iax and 50\% sub-Ch SNe Ia (green short-dashed lines).
The blue long-dashed, cyan dotted, and magenta dot-dashed lines are for dSph galaxies with 100\% SN Iax, 100\% sub-Ch SNe Ia, and equal mix of the two, respectively.
See \citet[black]{kob11agb} and \citet[color]{ven12} for the observational data sources of the solar neighborhood and dSphs, respectively.}
\end{figure*}

For stars in the mass range $\sim 8-10 M_\odot$ (at $Z=0.02\equiv Z_\odot$), electrons are partially degenerate in a C+O core.
In $\sim 8-9 M_\odot$ stars,
neutrino cooling and contraction leads to off-center ignition of C flame, which moves inward all the way to the center as a result of heat conduction, while $\sim 9-10 M_\odot$ stars undergo central carbon ignition.
For both cases, a strongly degenerate O+Ne+Mg core was formed (O+Ne dominant, but Mg is essential for electron capture).
If Ne burning is not ignited \citep{nom84}, or if off-center Ne burning does not propagate to the center \citep{jon14}, 
such a core eventually undergoes an electron-capture-induced collapse, which does not produce significant amounts of iron \citep{nom82c}.
If the stellar envelope is lost by winds or binary interaction, an O+Ne+Mg WD may be formed.

Recently, at $\ltsim 9 M_\odot$, \citet{che14} suggested that if the convective undershooting of
the C-burning layer is large enough, the carbon fraction below the flame is largely
reduced, 
and showed a possibility of hybrid C+O+Ne WDs with $\gtsim$ $1.1M_\odot$ \citep[see also][]{den14}.
This depends on
convective boundary mixing and mass loss in binary systems.
If such a hybrid WD is in a close binary, the WD mass can reach the Ch-mass with a small amount of accretion from various masses of companion stars.
\citet{men14} connected this system to SN Iax \citep[e.g.,][]{fol13} and estimated the rate (only for MS companions), which could be 1-8\% of the overall SN Ia rate.
\citet{kro13} showed that deflagration of a Ch-mass (C+O) WD model can match the observed light curve and spectra for 2002cx-like objects, the bright end of SNe Iax.

In our single degenerate scenario, WD winds play an essential role in increasing the parameter space of SN Ia progenitors.
\citet{kob98} showed the metallicity effect of the WD winds as a function of WD mass, and the metallicity limit is [Fe/H] $= -1.1$ for normal SNe Ia from $\sim 1 M_\odot$ WDs.
WDs more massive than $1.2M_\odot$ were supposed to collapse to neutron stars and not to explode as normal SNe Ia \citep{nom91}.
However, if it is possible to form and explode $\sim 1.3 M_\odot$ WDs, the metallicity limit becomes as low as [Fe/H] $= -2.5$ (Fig.\ref{fig:iax}).
In any case, O+Ne+Mg WDs will form at $> 1.3 M_\odot$.

In this second channel, we assume that
a hybrid WD undergoes central carbon deflagration at a similar $\dot{M}$ as for normal SNe Ia when it reaches $1.38 M_\odot$ \citep{nom82a}, but the carbon deflagration is quenched when it reaches the outer O+Ne layer.
The progenitor WD is only partially burnt and ejected, and therefore the $^{56}$Ni production is small.
These explosions are fainter than SNe Ia, and may correspond to SNe Iax.
The WD should be massive, $M_{\rm WD,0} \sim 1.1-1.3 M_\odot$, and thus the primary mass ranges are set above the ranges for normal SNe Ia (Table 1).
The secondary mass ranges are calculated for $M_{\rm WD,0}=1.2M_\odot$.

The ejecta mass and iron production are lower than for normal SNe Ia.
There will not be much C and O ejected, and thus [$\alpha$/Fe] is very small, which is different from double degenerate systems \citep{rop12}.
On the other hand, [Mn/Fe] should be rather high because of the deflagration, and there will be no strong metallicity effect on [Mn/Fe] as Mn is predominantly synthesized in NSE.
In our GCE, we adopt the nucleosynthesis yields of the N5def model in \citet{fin14}; $M({\rm Fe})=0.193M_\odot, M({\rm O})=0.060M_\odot, M({\rm Mn})=3.67 \times 10^{-3}M_\odot$, which give [O/Fe] $=-1.42$ and [Mn/Fe] $=0.22$.
As noted above, these yields are for the bright end of SNe Iax, but similar results are obtained even if we halve the SN Iax rate in our GCE.

\section{Chemical Evolution Model}

\begin{figure*}[t]
\center
\vspace*{1mm}
\includegraphics[width=15cm]{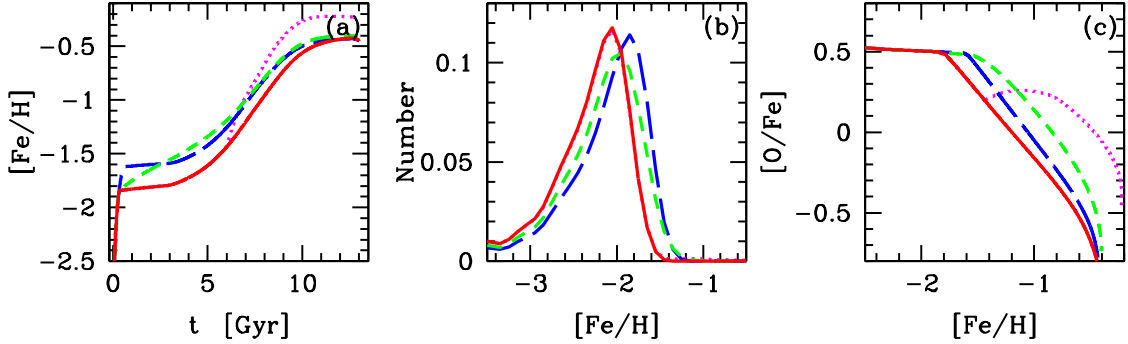}
\caption{\label{fig:dsph}
The age-metallicity relations (panel a), the MDF (b), and the [O/Fe]-[Fe/H] relations (c) for the models of dSph galaxies both with SN Iax and sub-Ch SNe Ia.
The adopted parameters are [$t_{\rm gw}$/Gyr, $\tau_2$/Gyr] $=$ [0.3, 1000] (solid lines), [0.3, 100] (short-dashed lines), [0.5, 1000] (long-dashed lines), and $t_2=6$Gyr for the model with a secondary star burst (dotted line).
}
\end{figure*}

The equations of our chemical evolution code are described in \citet{kob00}.
This is a so-called one-zone model, which assumes instantaneous mixing of the system.
The nucleosynthesis yields of core-collapse supernovae are taken from \citet{kob06}, with a hypernova fraction of 0.5 at $M \ge20M_\odot$.
The contribution from AGB stars is also included, and Kroupa (2008)'s IMF is adopted as in \citet{kob11agb}. 
The rates of sub-Ch SNe Ia and SNe Iax are calculated with Equation 12 of \citet{kob00}, which is the same as Equation 2 of \citet{kob09}.
In both cases, the binary parameters are the same as for normal SNe Ia; at $Z=0.004$, 2.3\% of the progenitor WDs (from $2.8-7.3M_\odot$) eventually explode, in both MS and RG systems.
We assume that neither the double degenerate channel nor double detonations accreting from He-WDs or He-burning stars contribute to the chemical enrichment of galaxies; otherwise, it is not possible to reproduce the elemental abundances of the solar neighborhood \citep{kob98,kob09}.

Figure \ref{fig:xfe} shows the evolution of elemental abundance ratios in the solar neighborhood and dSph galaxies.
For the solar neighborhood, the star formation history is determined to match the observed metallicity distribution function (MDF) of stars \citep[see Fig.12 of][]{kob11agb}, and consists of slow inflow (timescale of 5 Gyr) and star formation (timescale of 4.7 Gyr).
Because of the very small number of metal-poor stars, the contribution from SNe Iax and sub-Ch SNe Ia is negligible in the solar neighborhood. 
The difference between the solid and short-dashed lines is almost invisible in the [O/Fe]-[Fe/H] relation, which shows the plateau from core-collapse supernovae at low metallicities and the decreasing trend of [O/Fe] from [Fe/H] $\sim -1$ to $\sim 0$ due to the delayed enrichment of normal SNe Ia.

For dSph galaxies, the star formation history is uncertain, and should vary among galaxies. Because of the low star formation rate overall, the instantaneous mixing approximation may not be valid. Therefore, a large scatter around the model lines can be expected.
Figure \ref{fig:dsph} shows the age-metallicity relations, MDFs, and [O/Fe]-[Fe/H] relations of the models of dSph galaxies. The solid lines are the models used in Figure \ref{fig:xfe}.
In all of these models, the majority of stars formed in a short star burst (duration of $t_{\rm gw} = 0.3$Gyr) with a low efficiency (the timescale is $\tau_1=10$Gyr, which is even longer than in the solar neighborhood models).
Because of the small gravitational potential of the system, the initial star burst causes a strong galactic wind, which suppresses the following star formation by a factor of 100 (the timescale is $\tau_2=1000$Gyr)\footnote{SFR $= f_{\rm g}/\tau_1$ for $t\le t_{\rm gw}$ and $t\ge t_2$, and SFR $=f_{\rm g}/\tau_2$ for $t_{\rm gw} < t < t_2$, where $f_{\rm g}$ is the gas fraction.}, in addition to producing a continuous outflow proportional to the star formation rate.
Inflow is not included, and the initial gas fraction is set to be 1.
The iron abundance reaches only up to $\sim -2$ at the first burst, and the MDF peaks at [Fe/H] $\sim -2$, which is consistent with observations \citep{sta10}.
After the initial star burst, the metallicity gradually increases with SNe Iax and sub-Ch SNe Ia.

The mean stellar metallicity of galaxies and the metallicity of the [$\alpha$/Fe] knee depend on the duration and efficiency of star formation.
With $t_{\rm gw}=0.5$Gyr (long-dashed lines), the metallicity reaches up to [Fe/H] $\sim -1.5$ by the initial star burst, and with $\tau_2=100$Gyr (short-dashed lines), the metallicity linearly increases at [Fe/H] $\gtsim -2$ after the initial star burst (Fig.\ref{fig:dsph}a). Both result in a higher peak metallicity of MDFs (Fig.\ref{fig:dsph}b).
In the [O/Fe]-[Fe/H] relations (Fig.\ref{fig:dsph}c),
with $t_{\rm gw}=0.5$Gyr or $\tau_2=100$Gyr, the knee is located at a higher metallicity.
If there is a secondary star burst (at $t_2$, dotted lines), even with a small enhancement of star formation, [Fe/H] and [O/Fe] show a rapid increase.
These strong parameter dependencies may explain the observed large scatter among stars in a dSph and the variation among dSphs.

As shown in Figure \ref{fig:xfe}, both for the SNe Iax and sub-Ch SNe Ia models, the [$\alpha$/Fe] knee appears at a much lower metallicity than in the solar neighborhood, which is more consistent with the observations of stars in dSphs \citep[e.g.,][]{tol09}.
On the other hand, the SNe Iax and sub-Ch SNe Ia models show a notable difference in the [Mn/Fe] ratios.
If SNe Iax are the result of deflagrations, a significant amount of Mn should be produced at low metallicities, and [Mn/Fe] shows a rapid increase from the the [$\alpha$/Fe] knee, even assuming 10\% of the SNe Iax rate.
If sub-Ch SNe Ia produce much less Mn at low metallicities, [Mn/Fe] shows a decrease there.
Therefore, a model in which both SNe Iax and sub-Ch SNe Ia almost equally contribute can give no [Mn/Fe] evolution for a wide range of metallicity.
This is consistent with the observed [Mn/Fe] ratios in dSphs \citep{mcw03,nor12}, although the observational data plotted in Fig.\ref{fig:xfe} show a large scatter.
Further observational data are required to constrain the contributions from these subclasses of thermonuclear supernovae.

\section{Conclusions}

Subclasses of thermonuclear explosions, such as SNe Iax and sub-Ch mass SNe Ia, can be an important enrichment source in dSph galaxies in the Local Group.
We have presented chemical evolution models with the two possible progenitor channels.
Due to the metallicity dependence on the mass ranges of primary and secondary stars, the event rate is higher for lower metallicity environments, but the contribution is negligible in the solar neighborhood.
In dSph galaxies, the decrease of [$\alpha$/Fe] ratios at [Fe/H] $\sim -2$ to $-1.5$ can be produced depending on the star formation history.
SNe Iax give high [Mn/Fe], while sub-Ch-mass SNe Ia give low [Mn/Fe], and thus a model including a mix of the two is favoured by the available observations.
Further calculations of stellar evolution, explosion, nucleosynthesis, and binary population synthesis with various metallicities are required to constrain their contributions in dSph galaxies.

\acknowledgments
We thank K.Venn, A.Karakas, S.Jones, R.Scalzo, I.Seitenzahl, A.Ruiter., and A.Bunker for fruitful discussion.
This work has been supported in part by the WPI Initiative, MEXT, Japan, and by the Grants-in-Aid for Scientific Research of the JSPS (23224004, 23540262, 24540227, and 26400222).


\begin{thebibliography}{}

\bibitem[Anders \& Grevesse(1989)]{and89}
Anders, E., \& Grevesse, N. 1989, Geochimica et Cosmochimica Acta, 53, 197

\bibitem[Carretta et al.(2009)]{car09}
Carretta, E., Bragaglia, A., Gratton, R., D'Orazi, V., \& Lucatello, S. 2009, \aap, 508, 695

\bibitem[Cayrel et al.(2004)]{cay04}
Cayrel, R. et al. 2004, \aap, 416, 1117

\bibitem[Chen et al.(2014)]{che14}
Chen, M. C., Herwig, F., Denissenkov, P. A., \& Paxton, B. 2014, \mnras, 440, 1274

\bibitem[Denissenkov et al.(2015)]{den14}
Denissenkov, P., et al. 2015, MNRAS, 447, 2696

\bibitem[Fink et al.(2010)]{fin10}
Fink, M., et al. 2010, \aap, 514, A53

\bibitem[Fink et al.(2014)]{fin14}
Fink, M., et al. 2014, \mnras, 438, 1762

\bibitem[Foley et al.(2013)]{fol13}
Foley, R. J. et al. 2013, \apj, 767, 57

\bibitem[Gratton(1989)]{gra89}
Gratton, R. G. 1989, \aap, 211, 41

\bibitem[Hachisu \& Kato(2001)]{hac01}
Hachisu, I. \& Kato, M. 2001, \apj, 558, 323

\bibitem[Hachisu, Kato \& Nomoto(1996)]{hac96} 
Hachisu, I., Kato, M., \& Nomoto, K. 1996, \apj, 470, L97

\bibitem[Hachisu et al. (2012)]{hac12} 
Hachisu, I., Kato, M., Saio, H., \& Nomoto, K. 2012, \apj, 744, 69

\bibitem[Hillebrandt \& Niemeyer(2000)]{hil00}
Hillebrandt, W. \& Niemeyer, J. C. 2000, \araa, 38, 191

\bibitem[H\"{o}flich \& Khokhlov(1996)]{hof96}
H\"{o}flich, P., \& Khokhlov, A. 1996, \apj, 457, 500

\bibitem[Iben \& Tutukov(1991)]{ibe91}
Iben I., Jr. \& Tutukov A. V., 1991, \apj, 370, 615

\bibitem[Jones et al.(2014)]{jon14}
Jones, S. Hirschi, R., \& Nomoto, K. 2014, \apj, 797, 83

\bibitem[Kawai, Saio \& Nomoto(1987)]{kaw87}
Kawai, Y., Saio. H., \& Nomoto, K. 1987, \apj, 315, 229

\bibitem[Kobayashi et al.(2011)]{kob11agb} 
Kobayashi, C., Karakas, I. A., \& Umeda, H. 2011, \mnras, 414, 3231

\bibitem[Kobayashi \& Nomoto(2009)]{kob09}
Kobayashi, C., \& Nomoto, K. 2009, \apj, 707, 1466

\bibitem[Kobayashi et al.(2000)]{kob00} 
Kobayashi, C., Tsujimoto, T., \& Nomoto, K. 2000, \apj, 539, 26

\bibitem[Kobayashi et al.(1998)]{kob98}
Kobayashi, C., Tsujimoto, T., Nomoto, K., Hachisu, I, \& Kato, M. 1998, 
\apj, 503, L155

\bibitem[Kobayashi et al.(2006)]{kob06} 
Kobayashi, C., Umeda, H., Nomoto, K., Tominaga, N., \& Ohkubo, T. 2006, \apj, 653, 1145

\bibitem[Koch et al.(2008)]{koc08}
Koch, A., McWilliam, A., Grebel, E. K., Zucker, D. B., \& Belokurov, V. 2008, \apj, 688, L13

\bibitem[Kraft et al.(1997)]{kra97} 
Kraft, R. P., Sneden, C., Smith, G. H., Shetrone, M. D., Langer, G. E., Pilachowski, C. A. 1997, \aj, 113, 279

\bibitem[Kromer et al.(2010)]{kro10}
Kromer, M., et al. 2010, \apj, 719, 1067

\bibitem[Kromer et al.(2013)]{kro13}
Kromer, M., et al. 2013, \mnras, 429, 2287

\bibitem[Kroupa(2008)]{kro08}
Kroupa, P. 2008, ASP Conference Series, 390, 3

\bibitem[Li et al.(2011)]{li11}
Li, W. et al., 2011, \mnras, 412, 1441

\bibitem[Maoz, Mannucci \& Nelemans(2014)]{mao14}
Maoz, D., Mannucci, F., \& Nelemans, G. 2014, \araa, 52, 107

\bibitem[McWilliam, Rich \& Smecker-Hane(2003)]{mcw03}
McWilliam, A., Rich, R. M., \& Smecker-Hane, T. A. 2003, \apj, 592, 21

\bibitem[Meng \& Podsiadlowski(2014)]{men14}
Meng, X. \& Podsiadlowski, P. 2014, \apj, 789, L45

\bibitem[Nomoto(1982a)]{nom82a}
Nomoto, K. 1982a, \apj, 253, 798

\bibitem[Nomoto(1982b)]{nom82}
Nomoto, K. 1982b, \apj, 257, 780

\bibitem[Nomoto(1984)]{nom84}
Nomoto, K. 1984, \apj, 277, 791

\bibitem[Nomoto et al.(1997)]{nom97}
Nomoto, K., et al. 1997, Nuclear Physics, A621, 467c

\bibitem[Nomoto et al.(2013)]{nom13}
Nomoto, K., Kobayashi, C., \& Tominaga, N. 2013, \araa, 51, 457

\bibitem[Nomoto \& Kondo(1991)]{nom91}
Nomoto, K., \& Kondo, Y. 1991, \apj, 367, L19

\bibitem[Nomoto et al.(1982)]{nom82c}
Nomoto, K., Sparks, W. M., Fesen, R. A., et al. 1982, Nature, 299, 803 

\bibitem[North et al.(2012)]{nor12}
North, P. et al., 2012, \aap, 541, 45

\bibitem[Nugent et al.(1997)]{nug97}
Nugent, P., Baron, E., Branch, D., Fisher, A., Hauschildt, P. H., 1997, \apj, 485, 812

\bibitem[Romano, Cescutti, \& Matteucci(2011)]{rom11}
Romano, D., Cescutti, G., \& Matteucci, F. 2011, \mnras, 418, 696

\bibitem[R\"opke et al.(2012)]{rop12}
R\"opke, F. K., et al. 2012, \apj, 750, L19

\bibitem[Ruiter et al.(2014)]{rui14}
Ruiter, A. J., et al. 2014, \mnras, 440L, 101

\bibitem[Scalzo et al.(2014)]{sca14}
Scalzo, R. A., Ruiter, A. J., \& Sin, S. A. 2014, \mnras, 445, 2535

\bibitem[Seitenzahl et al.(2013)]{sei13}
Seitenzahl, I. R., et al. 2013, \aap, 559, L5S

\bibitem[Seitenzahl et al.(2015)]{sei15}
Seitenzahl, I. R., et al. 2015, \mnras, 447, 1484

\bibitem[Shen \& Bildsten(2009)]{she09}
Shen K. J. \& Bildsten L., 2009, \apj, 699, 1365

\bibitem[Shigeyama et al.(1992)]{shi92}
Shigeyama, T., et al. 1992, \apj, 386, L13

\bibitem[Sim et al.(2010)]{sim10}
Sim, S. A., Ro\"epke, F. K., Hillebrandt, W., Kromer, M., Pakmor, R., Fink, M., Ruiter, A. J., Seitenzahl, I. R., 2010, \apj, 714, L52

\bibitem[Starkenburg et al.(2010)]{sta10}
Starkenburg, E. et al. 2010, \aap, 513, 34

\bibitem[Tolstoy et al.(2009)]{tol09}
Tolstoy, E., Hill, V., \& Tosi, M. 2009, \araa, 47, 371

\bibitem[Tolstoy et al.(2003)]{tol03}
Tolstoy, E., Venn, K. A., Shetrone, M., Primas, F., Hill, V.,
Kaufer, A., \& Szeifert, T. 2003, \aj, 125, 707

\bibitem[Umeda et al.(1999)]{ume99}
Umeda, H., Nomoto, K., Yamaoka, H., \& Wanajo, S. 1999, \apj, 513, 861

\bibitem[Venn et al.(2012)]{ven12}
Venn, K. A. et al. 2012, \apj, 751, 102

\bibitem[Woosley \& Kasen(2011)]{woo11}	
Woosley, S. E. \& Kasen, D 2011, \apj, 734, 38

\bibitem[Yamaguchi et al.(2015)]{yam15}
Yamaguchi, H. et al. \apj, 801, L31

\bibitem[Yungelson et al.(1995)]{yun95}
Yungelson, L., Livio, M., Tutukov, A., Kenyon, S. J., 1995, apj, 447, 656 

\end{thebibliography}
\end{document}